\documentclass[10pt,aps,prd,twocolumn,superscriptaddress,nofootinbib,floatfix]{revtex4-2}

\usepackage{placeins}
\usepackage{amsmath,amssymb}
\usepackage{graphicx}
\usepackage{xcolor}
\usepackage{multirow}
\usepackage{array}
\usepackage{ragged2e}
\usepackage[normalem]{ulem}

\usepackage[colorlinks=true,linkcolor=red,citecolor=green,urlcolor=blue]{hyperref}

\newlength{\colA}
\newlength{\colB}
\newlength{\colC}

\begin{document}

\title{Fermion Mixing Matrices and the Exceptional Jordan Algebra}

\author{Bishnu Gupta Teli}
\email{ph24c802@smail.iitm.ac.in}
\affiliation{Department of Physics, Indian Institute of Technology Madras, Chennai 600036, India}

\author{Tejinder P. Singh}
\email{tpsingh@tifr.res.in}
\affiliation{Tata Institute of Fundamental Research, Homi Bhabha Road, Mumbai 400005, India}

\date{\today}

\begin{abstract}
We extend the exceptional-Jordan spectral framework for fermion mass hierarchies to the problem of quark and lepton mixing. Following the companion mass paper~\cite{Teli:2026jgr}, each fermion sector is associated with a Hermitian element of $J_3(\mathbb{O}_{\mathbb{C}})$, where adjacent square-root mass ratios are obtained from cubic ladders in $\mathrm{Sym}^3(\mathbf 3)$. Here, these ratios are used as inputs to an adjacent-edge lift from spectral hierarchy data to two-generation mixing angles. The lift is derived from a Fritzsch-type two-state texture~\cite{Fritzsch:1977za, Fritzsch:1979zq} and should be regarded as an effective bridge ansatz rather than a theorem of the Jordan spectrum alone. The exact CP-transport input is supplied by the companion CP Letter~\cite{GuptaTeli:2026aqf}. In the quark sector, the octonionic ladder operator $\alpha_2$ generates a real local rotor in the $(e_1,e_3)$ plane, and the up- and down-sector local Cabibbo-edge amplitudes are complex conjugates, giving the exact local law $\phi_{12}=-2\chi$. This is a transport-level Cabibbo-rung phase law, not by itself a prediction of the standard CKM Dirac phase. With the fitted companion mass ratios, the minimal two-angle extraction from the measured $|V_{us}|$ gives an effective Cabibbo-block phase $\phi_{12}\simeq 105.7^\circ$; this number is a bridge diagnostic, while the balanced octonionic rotor remains the distinguished quadrature reference point. The $(2,3)$ sector requires a phenomenological normalization $\kappa_{23}\simeq0.56$, and the direct $(1,3)$ element remains a long-edge bridge problem. Thus, the present CKM construction is not yet parameter-free: its derived content is the local rung phase structure, and its open content is the full weak-to-mass bridge. In the lepton sector, charged-lepton edge angles are small compared with the observed PMNS angles, so the large mixing must come from the neutral-sector diagonalizer, the Higgs bridge, or both. The neutrino Jordan spectrum by itself does not fix \(U_{\nu, L}\): in the symmetric branch, two eigenvalue magnitudes are degenerate or quasi-degenerate, leaving a neutral-sector basis freedom. At the level of the general mixing analysis, we do not assume a Dirac or Majorana interpretation or select a neutrino mass ordering. The CP Letter also discusses a separate minimal Majorana branch in which the leading spectrum \((-\delta_\nu,0,+\delta_\nu)\), together with a perturbative-lift assumption, gives inverted ordering and \(m_{\beta\beta}\simeq19\,{\rm meV}\). We quote that branch as an additional specialization, not as an input to the adjacent-edge lift. The transport theorem used here is conditional: if the final real-linear transport entering the lepton amplitudes does not mix the scalar line with the corresponding local lepton flavor plane, and if the neutral-sector bridge is also real in this sense, then all PMNS amplitudes are real up to removable phases and \(J_\ell=0\), \(\delta^\ell_{\rm CP}\in\{0,\pi\}\). A leptonic CP phase, therefore, locates scalar--flavor mixing, an intrinsically complex bridge, or neutral-sector bridge structure outside the safe real-linear class.
\end{abstract}

\maketitle

\section{Introduction}
The Standard Model describes flavor mixing through two unitary matrices. The Cabibbo-Kobayashi-Maskawa (CKM) matrix relates the weak and mass bases of quarks~\cite{Cabibbo1963, KobayashiMaskawa1973}, while the Pontecorvo-Maki-Nakagawa-Sakata (PMNS) matrix does the same for leptons~\cite{MakiNakagawaSakata1962, Pontecorvo1968}. Their entries are accurately measured~\cite{ParticleDataGroup:2024cfk, Esteban:2024fit, NuFIT6.1}, but the Standard Model treats the three mixing angles and the CP-violating phase in each sector as phenomenological Yukawa data. The origin of these numbers remains part of the flavor problem.

An early exploration of extracting CKM parameters in this algebraic context was carried out by Patel and Singh~\cite{Patel:2023qtw} using top-down state superpositions. Subsequently, a rigid exceptional-Jordan framework formulated by Singh~\cite{Singh:2025xxv} established closed-form, parameter-free square-root mass ratios from minimal cubic ladders with a fixed geometric spread $\delta^2=3/8$, following earlier exceptional-Jordan mass-ratio work~\cite{Bhatt:2021adg}. To optimize these relations against low-energy data, the recent companion paper~\cite{Teli:2026jgr} developed a broader spectral organization of fermion mass hierarchies by treating those rigid constraints as a baseline point within a constrained spectral moduli space of $J_3(\mathbb{O}_\mathbb{C})$. In that framework, one generation of Standard Model states is first realized inside $\mathbb{C}\otimes\mathbb{O}$ following the octonionic Clifford construction of Furey~\cite{Furey:2010fm, Furey:2015yxg,furey_2018_plb}, which builds on the older division-algebraic Standard-Model program~\cite{GunaydinGursey1973, Dixon1994}. Three generations are then associated with a cyclic triple of inequivalent octonionic complex structures. For each fermion species $f\in\{u,d,e,\nu\}$, these three generation states are embedded into a Hermitian element $A_f\in J_3(\mathbb{O}_\mathbb{C})$, the exceptional Jordan algebra introduced in the quantum-mechanical setting in Ref.~\cite{jordan_von_neumann_wigner_1934}. The Jordan eigenvalues give ordered spectral scales, and cubic ladders in $\mathrm{Sym}^3(\mathbf 3)$ organize the adjacent mass ratios.

The purpose of the present paper is to study how much of the mixing structure can be extracted from the same spectral data. The main idea is simple. If neighboring edges of the cubic ladder encode adjacent mass ratios, then the same edges should also control the size of the corresponding weak-basis rotations. This motivates an adjacent-edge lift from square-root mass ratios to two-generation mixing angles. In contrast with the spectral mass-ratio formulae of the companion paper, however, this lift is not a direct invariant statement about $J_3(\mathbb{O}_{\mathbb{C}})$. It is an effective bridge ansatz. Its status is therefore closer to a texture prescription than to a closed-form Jordan spectral theorem.

A companion CP Letter~\cite{GuptaTeli:2026aqf} supplies the transport-level phase theorem used below. The notation in that Letter is a local Clifford chart adapted to the Cabibbo rung, whereas the mass-hierarchy paper uses the spectral-family chart $\{e_6,e_3,e_5\}$ for the Jordan embedding. These charts should not be identified component by component. The invariant statements are instead chart-covariant: once the bridge identifies an adjacent quark edge with the local Cabibbo rung, the local rung supplies the exact phase law $\phi_{12}=-2\chi$; and once the final lepton bridge is represented by a safe real-linear transport, the lepton amplitudes are real up to removable phases.

This separation of roles is essential. In the quark sector, the CP Letter proves the conjugate-amplitude relation and the one-rung law $\phi_{12}=-2\chi$. It does not, by itself, extract the standard CKM Dirac phase. In the present paper, we embed this local result inside the adjacent-edge mixing ansatz. The numerical value $\phi_{12}\simeq 105.7^\circ$ obtained below from the measured Cabibbo magnitude is therefore an effective Cabibbo-block parameter of the bridge ansatz, not an independent CP datum and not a prediction of $\delta^q_{\rm CP}$. The remaining CKM sectors, especially the $(2,3)$ normalization and the direct $(1,3)$ long edge, still require dynamical bridge input. This value should not be conflated with the near-quadrature tilt number quoted in the CP Letter. There, the earlier balanced-down plus up-tilt construction gives \(\chi_{\rm eff}\simeq -32^\circ\), equivalently \(\phi_{12}\simeq 64^\circ\), whereas the present adjacent-edge lift gives \(\phi_{12}\simeq105.7^\circ\). The reconciliation is explicitly given in Sec.~\ref{subsec:ckm-phase-relation}: the two numbers correspond to different effective bridge assumptions.

The lepton sector requires the same bookkeeping. The mass-hierarchy paper treats the neutrino sector spectrally and leaves the neutral-sector diagonalizer undetermined at the level needed for PMNS mixing. The CP Letter adds a narrower minimal branch: if the symmetric spectrum is interpreted as a perturbatively lifted real Majorana spectrum, then the leading pattern $(-\delta_\nu,0,+\delta_\nu)$ gives inverted ordering and \(m_{\beta\beta}\simeq19\,{\rm meV}\). The present mixing paper does not use that extra minimality assumption to construct PMNS angles. Its default statement is weaker and more general: the neutrino spectrum alone does not determine \(U_{\nu, L}\), while the CP theorem constrains the final bridge. If the bridge is safe, real-linear, leptonic Dirac CP is conserved; if experiment establishes a nonzero \(\delta^\ell_{\rm CP}\), the result identifies scalar--flavor mixing, an intrinsically complex bridge, or neutral-sector bridge dynamics outside the safe class.

The paper is organized as follows. Sec.~\ref{sec:review} reviews the parts of the mass hierarchy framework needed for mixing. Sec.~\ref{sec:lift} introduces the adjacent-edge lift and establishes its effective status. Sec.~\ref{sec:quarks} analyzes quark mixing and derives the Cabibbo-rung phase law, while Sec.~\ref{sec:leptons} analyzes lepton mixing and the conditional reality theorem. Sec.~\ref{sec:phenomenology} then gives a phenomenological assessment and status table. We conclude in Sec.~\ref{sec:conclusion}, including a dynamical outlook based on trace dynamics.

\section{Review of the Spectral Mass Hierarchy}\label{sec:review}
Only the ingredients needed for mixing are recalled here. The detailed spectral fit and the discussion of the constrained Jordan moduli space are given in the companion mass paper~\cite{Teli:2026jgr}. The construction is not a conventional grand-unified model. The octonionic Clifford construction provides gauge-representation data, while the spectrum of Hermitian Jordan elements organizes inter-generation hierarchy data.

\subsection{Octonionic States and Generation Labels}
Let $\mathbb{O}$ be the real octonions with basis $\{1,e_1,\ldots,e_7\}$ and $e_a^2=-1$ for $a=1,\ldots,7$~\cite{Baez2002,Dixon1994,schafer_nonassociative}. Complexification gives $\mathbb{O}_{\mathbb{C}}=\mathbb{C}\otimes\mathbb{O}$, with the ordinary complex unit $i$ commuting with the octonionic units. For each imaginary octonionic direction $e_a$ define the left-multiplication operator~\cite{Teli:2026jgr, Furey:2015yxg}
\begin{equation}
    \mathcal{O}_a=\frac{i}{2}L_{e_a}, \qquad L_{e_a}v=e_a v.
\end{equation}
Since $L_{e_a}^2=-1$, the eigenvalues of $\mathcal{O}_a$ are $\pm 1/2$. A convenient set of eigenstates is
\begin{align}
    \nu^{(a)}&=\frac{1+i e_a}{2},& e^{(a)}&=\frac{1-i e_a}{2},\label{eq:leptonstates}\\
    u_k^{(a)}&=\frac{e_{b_k}+i e_{c_k}}{2},& d_k^{(a)}&=\frac{-e_{c_k}+i e_{b_k}}{2},\label{eq:quarkstates}
\end{align}
where $(e_{b_k},e_{c_k},e_a)$ are the three oriented quaternionic lines through $e_a$ in the Fano plane. After suppressing color by fixing one representative, a cyclic triple of complex structures, for example, $(e_6, e_3, e_5)$, is used to label the three generations. We write the corresponding states as $\psi_f^{(6)}$, $\psi_f^{(3)}$, and $\psi_f^{(5)}$. Later, in the CP analysis, we also use a local Clifford chart adapted to the Cabibbo ladder operator. This local chart is not a replacement for the spectral-family triple used in the Jordan embedding. In particular, the local lepton plane $\Pi_\ell^{\rm loc}=\mathrm{span}_{\mathbb R}\{e_7,e_5,e_2\}$ should be read as the theorem plane in the CP chart. In a different family convention, the same condition is obtained by transporting both the states and the plane. The physical criterion is not the literal label set $\{e_7,e_5,e_2\}$; it is whether the final bridge mixes $\mathbb C\cdot1$ with the relevant lepton flavor plane.

\paragraph*{Explicit chart transport.}
For clarity, one may exhibit an explicit signed Fano-plane automorphism relating the spectral-family chart used for the Jordan embedding to the local Clifford chart used in the CP theorem. One convenient choice is the real-linear map \(\mathcal T\in G_2=\mathrm{Aut}(\mathbb O)\), extended by \(\mathcal T(1)=1\), with
\begin{equation}\label{eq:explicit-chart-transport}
\begin{aligned}
\mathcal T(e_1)&=e_6,&
\mathcal T(e_2)&=e_3,&
\mathcal T(e_3)&=-e_5,&
\mathcal T(e_4)&=e_4,\\
\mathcal T(e_5)&=-e_2,&
\mathcal T(e_6)&=e_7,&
\mathcal T(e_7)&=e_1.&
\end{aligned}
\end{equation}
This map preserves the oriented Fano multiplication table, hence it is an octonion automorphism. It sends
\[
\mathrm{span}_{\mathbb R}\{e_6,e_3,e_5\}
\longmapsto
\mathrm{span}_{\mathbb R}\{e_7,e_5,e_2\},
\]
up to harmless signs. Thus, the spectral-family triple and the local lepton theorem plane are related by transport, not by componentwise identification. In particular, the fact that the symbol \(e_5\) appears in both triples does not mean that \(e_5\) is held fixed; in the transport above, \(\mathcal T(e_3)=-e_5\) and \(\mathcal T(e_5)=-e_2\). Since every element of \(G_2\) fixes the scalar line and preserves \(\Im\mathbb O\), \(\mathcal T\) is safe in the sense of Eq.~\eqref{eq:safe-condition}. Therefore, the lepton-reality condition is chart-covariant: in a transported chart, one replaces both the states and the lepton flavor plane by their \(\mathcal T\)-images.

\subsection{Hermitian Jordan Element}
For each fermion type $f\in\{u,d,e,\nu\}$, the three family states are placed in the off-diagonal slots of a Hermitian Jordan element
\begin{equation}
 A_f=
 \begin{pmatrix}
    r s_f & x_f & y_f^\dagger\\
    x_f^\dagger & r s_f & z_f\\
    y_f & z_f^\dagger & r s_f
 \end{pmatrix}
 \in J_3(\mathbb{O}_{\mathbb{C}}),
 \label{eq:JordanAf}
\end{equation}
with
\begin{equation}
    x_f=\psi_f^{(6)},\qquad
    y_f=\psi_f^{(3)},\qquad
    z_f=\psi_f^{(5)}.
\end{equation}
The family-center parameters are inherited from the octonionic charge assignment,
\begin{equation}
    s_d=1,\quad s_u=\frac{2}{3},
    \quad s_e=\frac{1}{3},\quad s_\nu=0,
\end{equation}
and $r$ is the relative normalization between diagonal and off-diagonal parts. The characteristic equation is
\begin{equation}\label{eq:charpoly}
    \lambda^3-\operatorname{tr}(A_f)\lambda^2+\sigma(A_f)\lambda-\det(A_f)=0.
\end{equation}
The exceptional Jordan eigenvalue problem and its characteristic invariants have been studied in detail in Refs.~\cite{DrayManogue1999, DuboisViolette2016,TodorovDuboisViolette2018}. For the symmetric elements used in the hierarchy construction,
\begin{align}
    \operatorname{tr} A_f&=3rs_f,\label{eq:tr}\\
    \sigma(A_f)&=3r^2s_f^2-\delta_f^2,\label{eq:sigma}\\
    \det(A_f)&=r^3s_f^3-rs_f\delta_f^2+2\tau_f,\label{eq:det}
\end{align}
where
\begin{align}
    \delta_f^2&=\operatorname{Sc}(x_f x_f^\dagger)+\operatorname{Sc}(y_f y_f^\dagger)+\operatorname{Sc}(z_f z_f^\dagger),\label{eq:delta}\\
    \tau_f&=\Re\bigl(y_f(x_fz_f)\bigr).\label{eq:tau}
\end{align}
The normalization condition enforces $\delta_f^2=3/8$ for all sectors~\cite{Teli:2026jgr}. The quark sectors have $\tau_u=\tau_d=0$, while the charged-lepton sector has an effective phase through $\tau_e=(1/64)\cos\Phi_e$. In the mass fit of Ref.~\cite{Teli:2026jgr},
\begin{equation}
    r=-0.987,\qquad p=0.987,
    \qquad \cos\Phi_e=-0.509.\label{eq:massfit}
\end{equation}

Ordering the positive magnitudes of the three roots gives
\begin{equation}
    a_f\le b_f\le c_f,
    \qquad \{a_f,b_f,c_f\}=\{|\lambda_1|,|\lambda_2|,|\lambda_3|\}_{\rm ordered} .
\end{equation}
These are the spectral scales used in the cubic ladder.

\subsection{Cubic Ladders and Adjacent Ratios}
The Jordan characteristic equation is cubic, and the hierarchy construction assigns generation amplitudes to monomials in the symmetric cubic representation $\mathrm{Sym}^3(\mathbf 3)$. The minimal monotone chains used in the companion paper are
\begin{align}
    \sqrt{m_d}&\propto [a_d^2b_d]^p,&
    \sqrt{m_s}&\propto [a_db_dc_d]^p,&
    \sqrt{m_b}&\propto [c_d^3]^p,\label{eq:downladder}\\
    \sqrt{m_u}&\propto [a_u^2b_u]^p,&
    \sqrt{m_c}&\propto [a_ub_uc_u]^p,&
    \sqrt{m_t}&\propto (b_u^2c_u)^p,\label{eq:upladder}\\
    \sqrt{m_e}&\propto [a_e^2c_e]^p,&
    \sqrt{m_\mu}&\propto [a_eb_ec_e]^p,&
    \sqrt{m_\tau}&\propto [b_e^3]^p.\label{eq:elladder}
\end{align}
The charged-lepton ladder is the Dynkin-reflected image of the down-quark ladder, with $b\leftrightarrow c$. The adjacent square-root mass ratios are therefore
\begin{align}
    r^d_{12}&\equiv \sqrt{\frac{m_d}{m_s}}=\biggl[\frac{a_d}{c_d}\biggr]^p,
    & r^d_{23}&\equiv \sqrt{\frac{m_s}{m_b}}=\biggl[\frac{a_db_d}{c_d^2}\biggr]^p,\label{eq:rd}\\
    r^u_{12}&\equiv \sqrt{\frac{m_u}{m_c}}=\biggl[\frac{a_u}{c_u}\biggr]^p,
    & r^u_{23}&\equiv \sqrt{\frac{m_c}{m_t}}=\biggl[\frac{a_u}{b_u}\biggr]^p,\label{eq:ru}\\
    r^e_{12}&\equiv \sqrt{\frac{m_e}{m_\mu}}=\biggl[\frac{a_e}{b_e}\biggr]^p,
    & r^e_{23}&\equiv \sqrt{\frac{m_\mu}{m_\tau}}=\biggl[\frac{a_ec_e}{b_e^2}\biggr]^p.\label{eq:re}
\end{align}
The numerical values used below are the fitted square-root hierarchy inputs of the companion mass-ratio paper, expressed in the adjacent-edge orientation. Since the companion table lists the corresponding inverse large ratios, we use their reciprocals:
\begin{align}
r^d_{12}
&\equiv \sqrt{\frac{m_d}{m_s}}
 \simeq 0.2388,
&
r^d_{23}
&\equiv \sqrt{\frac{m_s}{m_b}}
 \simeq 0.1483,
\nonumber\\
r^u_{12}
&\equiv \sqrt{\frac{m_u}{m_c}}
 \simeq 0.0377,
&
r^u_{23}
&\equiv \sqrt{\frac{m_c}{m_t}}
 \simeq 0.0722,
\nonumber\\
r^e_{12}
&\equiv \sqrt{\frac{m_e}{m_\mu}}
 \simeq 0.0658,
&
r^e_{23}
&\equiv \sqrt{\frac{m_\mu}{m_\tau}}\Bigg|_{\rm fit}
 \simeq 0.2388.
\label{eq:edge-ratios-numerical}
\end{align}
These fitted reciprocals are treated here as inputs to the mixing problem. The last entry should be read as the fitted spectral input inherited from the companion hierarchy fit, not as a fresh evaluation of the pole-mass ratio. Using charged-lepton pole masses gives \(\sqrt{m_\mu/m_\tau}\simeq0.2439\)~\cite{ParticleDataGroup:2024cfk}; the difference is about two percent and would shift the corresponding angle only from \(\theta^{(e)}_{23}\simeq13.4^\circ\) to \(\theta^{(e)}_{23}\simeq13.7^\circ\). We keep the fitted input in Eq.~\eqref{eq:edge-ratios-numerical} for uniformity with the quark-sector inputs.

\section{Adjacent-Edge Lift}\label{sec:lift}
Let $U_{f, L}$ be the left unitary matrix that diagonalizes the effective mass matrix of the fermion sector $f$. Then
\begin{equation}
    V_{\rm CKM}=U_{u,L}^\dagger U_{d,L},
    \qquad
    U_{\rm PMNS}=U_{e,L}^\dagger U_{\nu,L}. \label{eq:mixdefs}
\end{equation}
Here, the subscript \(L\) denotes the diagonalizer entering the charged-current weak-mixing amplitudes after the weak-to-mass bridge has been specified. It is standard CKM/PMNS notation only; it is not an additional left-handed/right-handed assumption in the spectral mass-hierarchy construction.
The Jordan spectrum fixes the mass-ratio data entering Eqs.~\eqref{eq:rd}--\eqref{eq:re}, but it does not by itself fix the complete matrices $U_{f, L}$. A bridge from spectra to eigenvectors is required. The adjacent-edge lift supplies a minimal such bridge.

Consider two generations with positive physical masses $m_i<m_j$. The two-state texture
\begin{equation}
    M_{ij}=\begin{pmatrix}
        0 & \sqrt{m_i m_j}\\
        \sqrt{m_i m_j} & m_j-m_i
    \end{pmatrix}\label{eq:fritzsch2}
\end{equation}
has eigenvalues $-m_i$ and $m_j$. The relative minus sign is unphysical for the mixing angle and can be absorbed by a diagonal rephasing of the corresponding mass eigenstate. If $R_{ij}(\theta)$ is the real rotation that diagonalizes this matrix, then
\begin{equation}
    \tan 2\theta_{ij}=\frac{2\sqrt{m_i m_j}}{m_j-m_i} .
\end{equation}
Writing $t=\tan\theta_{ij}$ gives
\begin{equation}
    \frac{2t}{1-t^2}=\frac{2\sqrt{m_i m_j}}{m_j-m_i},
\end{equation}
whose physical solution is exactly
\begin{equation}\label{eq:liftsqrt}
    \tan\theta_{ij}=\sqrt{\frac{m_i}{m_j}}.
\end{equation}
Thus, the square-root mass ratio is the natural edge angle of the two-state texture, in the lineage of the classic square-root hierarchy relations for the Cabibbo angle~\cite{GattoSartoriTonin1968, WilczekZee1977}.

Eq.~\eqref{eq:liftsqrt} is the adjacent-edge lift. Its role in this paper is not to prove that the Jordan spectrum uniquely determines the Yukawa eigenvectors. Rather, it gives the simplest local texture compatible with the classic square-root hierarchy relations~\cite{Fritzsch:1977za, Fritzsch:1979zq} already selected by the cubic ladders. Applying it to the full three-generation problem is a phenomenological ansatz:
\begin{equation}\label{eq:edgeansatz}
    t_{ij}^{(f)}\equiv \tan\theta_{ij}^{(f)}=r_{ij}^{(f)}
    \qquad (ij=12,23),
\end{equation}
with shorthand sine and cosine parameters defined by
\begin{equation}\label{eq:sincosfromtan}
    s_{ij}^{(f)}\equiv\frac{t_{ij}^{(f)}}{\sqrt{1+(t_{ij}^{(f)})^2}},
    \qquad
    c_{ij}^{(f)}\equiv\frac{1}{\sqrt{1+(t_{ij}^{(f)})^2}}.
\end{equation}

\paragraph*{Long Edges and the Limits of the Lift.} The lift in Eq.~\eqref{eq:edgeansatz} is an adjacent-edge prescription. It determines the rotations associated with the neighboring rungs $(1,2)$ and $(2,3)$ of the hierarchy. It does not automatically determine the direct long edge $(1,3)$. A naive product of adjacent ratios would give
\begin{equation}\label{eq:longproduct}
    r_{13}^{(f)}\sim r_{12}^{(f)}r_{23}^{(f)},
\end{equation}
which is not small enough to account for the observed $|V_{ub}|$ if both up and down long edges enter without additional suppression or phase structure. Consequently, the $(1,3)$ sector is treated below as a separate bridge problem. This separation is essential for keeping the predictive status of the construction explicit.

\section{Quark Mixing}\label{sec:quarks}

\subsection{CKM Rotations and the Cabibbo Block}
We write the left diagonalizers schematically as
\begin{equation}\label{eq:Ufrot}
    U_f=R_{23}^{(f)}R_{13}^{(f)}R_{12}^{(f)},
    \qquad f=u,d, 
\end{equation}
where phases may be inserted between the real rotations. To leading order in the $(1,2)$ block, and neglecting direct $(1,3)$ effects, the Cabibbo element has the form
\begin{equation}\label{eq:Vusleading}
    V_{us}=c_{12}^{(u)}s_{12}^{(d)}
    + e^{-i\phi_{12}}s_{12}^{(u)}c_{12}^{(d)},
\end{equation}
up to an unphysical rephasing convention. Therefore
\begin{align}\label{eq:Vusphase}
    |V_{us}|^2={}&
    \bigl(c_{12}^{(u)}s_{12}^{(d)}\bigr)^2
    +\bigl(s_{12}^{(u)}c_{12}^{(d)}\bigr)^2 \nonumber\\
    &+2c_{12}^{(u)}s_{12}^{(d)}s_{12}^{(u)}c_{12}^{(d)}\cos\phi_{12}.
\end{align}
Using
\[
t^{(u)}_{12}=0.0377,\qquad
t^{(d)}_{12}=0.2388,\qquad
|V_{us}|=0.22497,
\]
Eq.~\eqref{eq:Vusphase} gives
\begin{equation}
\cos\phi_{12}\simeq -0.271,
\qquad
\phi_{12}\simeq 105.7^\circ .
\label{eq:cabibbo-phase-fit}
\end{equation}
The small-angle version of the same extraction gives \(\phi_{12}\simeq 115.7^\circ\). We use the exact \(s=t/\sqrt{1+t^2}\), \(c=1/\sqrt{1+t^2}\) implementation as the quoted value. Equation~\eqref{eq:cabibbo-phase-fit}, however, is not a measurement of a CP phase and is not a derivation of the standard CKM phase. It is the value of the relative Cabibbo-block phase required by the minimal two-angle bridge ansatz after the fitted hierarchy ratios are inserted. The balanced octonionic rotor remains the natural quadrature reference point. The difference between the balanced value and Eq.~\eqref{eq:cabibbo-phase-fit} should be read as information about the still-uncomputed weak-to-mass bridge, not as a failure or success of the local transport theorem.

\subsection{Octonionic Origin of the Cabibbo Phase}
The phase in Eq.~\eqref{eq:Vusleading} is not arbitrary. It is the local Cabibbo-rung phase proved in the companion CP Letter~\cite{GuptaTeli:2026aqf}. We use the same fixed-$e_7$ local Clifford chart here because it is adapted to the ladder operator that realizes the adjacent Cabibbo edge. This chart is distinct from the spectral-family chart $\{e_6,e_3,e_5\}$ used to build the Jordan mass hierarchies; the bridge ansatz is what relates the adjacent spectral edge to this local transport calculation. In this local chart, the ladder operators may be written
\begin{align}\label{eq:alphas}
    \alpha_1=\frac{-\overleftarrow{e_5}+i \overleftarrow{e_4}}{2},
    \quad \alpha_2=\frac{-\overleftarrow{e_3}+i \overleftarrow{e_1}}{2},
    \quad \alpha_3=\frac{-\overleftarrow{e_6}+i \overleftarrow{e_2}}{2}.      
\end{align}
They satisfy
\begin{equation}\label{eq:clifford-relations}
    \{\alpha_i,\alpha_j\}=0,
    \qquad
    \{\alpha_i,\alpha_j^\dagger\}=\delta_{ij}.
\end{equation}
The adjoint is the combined complex and octonionic adjoint. Hence
\begin{equation}\label{eq:alpha2dagger}
    \alpha_2^\dagger=\frac{e_3+i e_1}{2}.
\end{equation}
The most general one-rung generator with orientation $\chi$ is
\begin{align}\label{eq:realrung}
    e^{i\chi}\alpha_2^\dagger-e^{-i\chi}\alpha_2
    &=\frac{1}{2}\left[e^{i\chi}(e_3+i e_1)-e^{-i\chi}(-e_3+i e_1)\right]
    \nonumber\\
    &=\cos\chi\,e_3-\sin\chi\,e_1\equiv g_\chi.
\end{align}
The important point is that $g_\chi$ contains no external complex unit $i$; it is a real octonionic imaginary unit satisfying $g_\chi^2=-1$ and lying in the plane $\operatorname{span}\{e_1,e_3\}$. Therefore
\begin{equation}\label{eq:rungrotor}
    U(\theta,\chi)=L_{\exp(\theta g_\chi)}
\end{equation}
is a real octonionic transport.

Let $K$ denote external complex conjugation, $K(i)=-i$, $K(e_a)=e_a$. The local up-type representatives on this Cabibbo edge and the corresponding anti-down representatives obey
\begin{equation}\label{eq:dukconj}
    \bar d_k=i K(u_k),\qquad k=1,2,
\end{equation}
up to irrelevant normalizations and rephasings. Since a real octonionic transport commutes with $K$, the transition amplitudes
\begin{equation}\label{eq:Audef}
    A_u=\langle u_2|U|u_1\rangle,
    \qquad
    A_d=\langle \bar d_2|U|\bar d_1\rangle
\end{equation}
are related by
\begin{equation}\label{eq:conjugationlaw}
    A_d=A_u^*.
\end{equation}
Here \(A_d\) is written in the anti-down representative natural to the octonionic ideal construction. Passing to the usual down-quark CKM convention only changes charge-conjugation and external rephasing conventions, and does not change the relative rung phase extracted from the conjugate amplitudes. A direct evaluation of the rotor gives
\begin{equation}\label{eq:AuAd}
    A_u=\mathcal{N}e^{-i\chi}\sin\theta,
    \qquad
    A_d=\mathcal{N}e^{+i\chi}\sin\theta,
    \qquad \mathcal{N}\in\mathbb{R}.
\end{equation}
Thus, the relative phase is
\begin{equation}\label{eq:phaseLaw}
    \phi_{12}=-2\chi.
\end{equation}
The balanced orientation $\chi=-\pi/4$ gives $\phi_{12}=\pi/2$. Within the two-angle bridge diagnostic of Eq.~\eqref{eq:Vusleading}, the effective Cabibbo-block value in Eq.~\eqref{eq:cabibbo-phase-fit} may be parametrized as
\begin{equation}\label{eq:chi-fit}
\chi_{12}^{\rm bridge}=-\frac{\phi_{12}}{2}\simeq -52.9^\circ .
\end{equation}
We write \(\chi_{12}^{\rm bridge}\) to avoid confusing this bridge parameter with a dynamically computed Yukawa orientation. The common derived statement is the conjugate-amplitude relation \(A_d=A_u^*\) and the local law \(\phi_{12}=-2\chi\). Identifying any such local block phase with the convention-independent CKM phase requires the full three-generation weak-to-mass bridge.

\subsection{Relation to the Standard CKM Phase}\label{subsec:ckm-phase-relation}
The relative phase \(\phi_{12}\) is a phase in a particular weak-basis construction. The standard CKM phase \(\delta_{\rm CKM}\) is a convention-independent parameter extracted, for example, from the Jarlskog invariant~\cite{Jarlskog1985}
\begin{equation}\label{eq:Jq}
    J_q=c_{12}c_{23}c_{13}^2s_{12}s_{23}s_{13}\sin\delta_{\rm CKM}.
\end{equation}
In a complete three-generation model, the relation between \(\phi_{12}\) and \(\delta_{\rm CKM}\) depends on the embedding of the \((2,3)\) and \((1,3)\) bridge sectors and on the associated rephasings. Therefore, in the fitted-moduli implementation used here, the phase extracted from \(|V_{us}|\) is not identified with \(\delta_{\rm CKM}\) as a separate theorem.

The numerical value in Eq.~\eqref{eq:cabibbo-phase-fit} should also be distinguished from the effective Cabibbo-block number quoted in the CP Letter. The CP Letter refers to the earlier tilt construction of Ref.~\cite{Singh:2025xxv}: a balanced-down reference rotor, amplitude-equivalent to \(\chi=-\pi/4\), is supplemented by a single real up-leg tilt \(\epsilon\simeq -26.1^\circ\). This gives
\[
\chi_{\rm eff}
=
-\frac12\left(\frac{\pi}{2}+\epsilon\right)
\simeq -32^\circ,
\qquad
\phi^{\rm tilt}_{12}=-2\chi_{\rm eff}\simeq 64^\circ .
\]
The present paper uses a different bridge. Both adjacent legs are fixed by the square-root-ratio lift,
\[
t^{(d)}_{12}=r^d_{12}\simeq0.2388,\qquad
t^{(u)}_{12}=r^u_{12}\simeq0.0377 .
\]
With the exact sine convention, this gives
\[
s^{(d)}_{12}
=
\frac{t^{(d)}_{12}}{\sqrt{1+(t^{(d)}_{12})^2}}
\simeq0.2323
>
|V_{us}|=0.22497 .
\]
Thus, the down-sector Cabibbo leg is already larger than the observed Cabibbo magnitude. In Eq.~\eqref{eq:Vusphase}, the up-sector leg must therefore subtract rather than add, forcing
\[
\cos\phi_{12}<0,\qquad \phi_{12}>\frac{\pi}{2}.
\]
This is why the adjacent-edge bridge gives \(\phi^{\rm edge}_{12}\simeq105.7^\circ\), on the opposite side of quadrature from the earlier tilt value. The two numbers are not competing predictions for \(\delta_{\rm CKM}\); they are diagnostics of two different weak-to-mass bridge assumptions. The common derived statement is the local octonionic law \(\phi_{12}=-2\chi\), together with the conjugate-amplitude relation \(A_d=A_u^*\).

The present paper, therefore, does not claim a numerical prediction for the standard CKM Dirac phase. The observed CKM phase motivates the search for an order-one octonionic orientation, but the comparison cannot be made before the full bridge is known. A parameter-free value of the standard phase requires a bridge calculation that fixes the \((2,3)\) normalization, the \((1,3)\) long edge, and the rephasing convention entering the Jarlskog invariant. The result established here is the source and form of the local quark phase, not its final embedding into the standard CKM parameterization.

\subsection{\texorpdfstring{The $\mathbf{(2,3)}$ Block}{The (2,3) Block}}
The adjacent-edge lift gives
\begin{equation}\label{eq:t23-corrected}
t^{(u)}_{23}=r^{(u)}_{23}\simeq 0.0722,
\qquad
t^{(d)}_{23}=r^{(d)}_{23}\simeq 0.1483.
\end{equation}
If the relative phase in this block is set to zero and no additional normalization is introduced, then
\begin{equation}\label{eq:vcb-zero-corrected}
|V_{cb}|_0 \simeq \left|s^{(d)}_{23}c^{(u)}_{23}-c^{(d)}_{23}s^{(u)}_{23}\right|\simeq 0.0751.
\end{equation}
The observed value is smaller than this estimate. We therefore introduce an effective normalization
\begin{equation}\label{eq:kappa23-def}
|V_{cb}|=\kappa_{23}\left|s^{(d)}_{23}c^{(u)}_{23}-c^{(d)}_{23}s^{(u)}_{23}\right|,
\end{equation}
which gives
\begin{equation}\label{eq:kappa23-corrected}
\kappa_{23}\simeq 0.56.
\end{equation}
This is a phenomenological parameter of the present implementation. It may encode the nonlocal nature of the \((2,3)\) bridge, the failure of independent two-state textures for a full three-state problem, or a phase structure associated with the other octonionic rungs. It is not yet derived from the exceptional Jordan spectrum.

\subsection{\texorpdfstring{The $\mathbf{(1,3)}$ Block}{The (1,3) Block}}
The direct element \(V_{ub}\) is more delicate. A literal adjacent-product estimate gives
\begin{equation}\label{eq:r13-corrected}
r^{(u)}_{13}\sim r^{(u)}_{12}r^{(u)}_{23}\simeq 2.7\times 10^{-3},
\qquad
r^{(d)}_{13}\sim r^{(d)}_{12}r^{(d)}_{23}\simeq 3.5\times 10^{-2}.
\end{equation}
The up-sector product alone is in the vicinity of the observed \(|V_{ub}|\) scale, whereas the down-sector product is an order of magnitude too large. Thus, the long-edge sector cannot be obtained by inserting both adjacent products as unsuppressed direct angles. The observed value, \(|V_{ub}|\simeq 3.7\times 10^{-3}\), requires a separate long-edge texture, suppression by the bridge operator, or destructive phase structure, especially in the down-sector contribution.

This point is conceptually useful. The cubic ladder gives adjacent hierarchy data, but a direct $(1,3)$ mixing matrix element is not simply another adjacent edge. It is a long-distance transition across the weight diagram. The present framework should therefore regard $V_{ub}$ as a constraint on the still-uncomputed bridge rather than as a parameter-free prediction of the adjacent-edge lift.

\section{Lepton Mixing}\label{sec:leptons}

\subsection{Charged-Lepton Rotations}
Applying the adjacent-edge lift to the charged-lepton ratios gives
\begin{equation}\label{eq:charged-lepton-t-corrected}
t^{(e)}_{12}\simeq 0.0658,
\qquad
t^{(e)}_{23}\simeq 0.2388 .
\end{equation}
A product estimate for the direct long edge gives
\begin{equation}\label{eq:charged-lepton-t13-corrected}
t^{(e)}_{13}\sim t^{(e)}_{12}t^{(e)}_{23}\simeq 1.57\times 10^{-2}.
\end{equation}
Thus
\begin{equation}\label{eq:charged-lepton-angles-corrected}
\theta^{(e)}_{12}\simeq 3.8^\circ,
\qquad
\theta^{(e)}_{23}\simeq 13.4^\circ,
\qquad
\theta^{(e)}_{13}\simeq 0.90^\circ .
\end{equation}
These angles are not large enough to account for the observed PMNS pattern on their own. In the present framework, the large lepton mixing must therefore come primarily from the neutral-sector diagonalizer \(U_{\nu, L}\), from the Higgs bridge, or from both.

\subsection{Neutrino Spectrum and Undetermined Diagonalizer}
For the neutrino Jordan element, the family center is $s_\nu=0$. The cubic invariant
\begin{equation}\label{eq:taunu}
    \tau_\nu=\frac{1}{64}\cos\Phi_\nu
\end{equation}
is not fixed by the charged-fermion mass fit. The eigenvalues can be written in trigonometric form as
\begin{align}\label{eq:nueigs}
    \lambda_k&=\frac{2\delta_\nu}{\sqrt{3}}
    \cos\left(\theta_\nu-\frac{2\pi k}{3}\right),\nonumber\\
    \cos(3\theta_\nu)&=\frac{3\sqrt{3}\tau_\nu}{\delta_\nu^3},
    \qquad k=0,1,2.
\end{align}
In the symmetric branch $\tau_\nu=0$, the relative spectrum is
\begin{equation}\label{eq:nusymmetric}
\lambda_\nu=(-\delta_\nu,0,+\delta_\nu),
\end{equation}
up to ordering.\footnote{Here \(\delta_\nu\) is the same universal Jordan spread used in Eq.~\eqref{eq:delta}: \(\delta_\nu^2=\delta_f^2=3/8\) for all sectors, so \(\delta_\nu=\sqrt{3/8}\simeq0.612\). The companion CP Letter uses the same relative spectrum \((-\delta_\nu,0,+\delta_\nu)\); any different dimensionless normalization assigned to the symbol \(\delta_\nu\) is a convention and is not an observable input in the present mixing analysis. The physical statements quoted below depend on the relative signs and on the later physical mass-scale assignment, not on this dimensionless normalization.} If neutrino masses are functions only of the magnitudes of these eigenvalues, the two nonzero states form a degenerate pair in magnitude. The mass eigenvalues alone, then, leave the basis within this degenerate two-dimensional subspace undetermined. Therefore, the spectrum by itself does not determine \(U_{\nu, L}\), and hence does not determine the PMNS angles.

At the level of the general mixing analysis, we do not impose a Dirac or Majorana interpretation of this neutral sector. Nor do we use the symmetric spectrum alone to select a mass ordering. This is the broad framework inherited from the mass paper: the spectral data organize the hierarchy, while the neutral-sector diagonalizer is part of the still-undetermined weak-to-mass bridge.

The companion CP Letter~\cite{GuptaTeli:2026aqf} studies an additional minimal specialization. If the opposite signs in the symmetric spectrum are interpreted as opposite Majorana parities of a real symmetric neutral mass matrix, and if the perturbations that generate the solar splitting are small enough not to reorder the leading spectrum, then the pair \((-\delta_\nu,+\delta_\nu)\) is assigned to the solar doublet and the zero eigenvalue to the light state. In that minimal branch,
\[
m_1\simeq m_2\simeq \sqrt{|\Delta m^2_{31}|},\qquad
m_3\simeq 0,
\]
so the ordering is inverted and
\[
m_{\beta\beta}\simeq m\,c_{13}^2|\cos 2\theta_{12}|
\simeq 19\,{\rm meV},\qquad
\Sigma m_\nu\simeq 0.10\,{\rm eV}.
\]
We quote this branch because it is part of the companion CP Letter, but we do not use it as an input to the adjacent-edge mixing lift. Thus, the present paper keeps two statements distinct: the general mixing framework leaves \(U_{\nu, L}\) and the ordering to the bridge, while the CP Letter's minimal Majorana branch gives a sharper, separately testable lepton-sector specialization.

\subsection{Lepton Flavor Plane and Safe Transports}\label{subsec:lepton-safe-transports}
Although the spectrum does not fix the lepton mixing angles, the CP Letter~\cite{GuptaTeli:2026aqf} proves a chart-covariant constraint on the final lepton transport. In the local Clifford chart used for that proof, the lepton plane is
\begin{equation}\label{eq:local-lepton-plane}
    \Pi_\ell^{\rm loc}=\mathrm{span}_{\mathbb R}\{e_7,e_5,e_2\}\subset\Im\mathbb O.
\end{equation}
This local plane is not the spectral-family triple $\{e_6,e_3,e_5\}$ used in the mass paper, and it should not be read as a new physical neutrino basis. Rather, it is the theorem plane selected by the local Clifford chart in which the transport statement is written. In any other cyclic generation convention, the same theorem is obtained by transporting the plane and the states together.

Let $\operatorname{C}:\mathbb O_{\mathbb C}\to\mathbb O_{\mathbb C}$ be the final complex-linear map with real coefficients in the basis $\{1,e_1,\ldots,e_7\}$. We say that $\operatorname{C}$ is safe if it does not mix the scalar line with the local lepton plane:
\begin{equation}\label{eq:safe-condition}
    [\operatorname{C}(1)]_{e_a}=0,\qquad [\operatorname{C}(e_a)]_0=0,\qquad e_a\in\Pi_\ell^{\rm loc}.
\end{equation}
Equivalently, in a transported chart, the same condition holds with $\Pi_\ell^{\rm loc}$ replaced by the corresponding transported lepton plane. Automorphisms in $G_2$ are safe because they fix $1$ and preserve $\Im\mathbb O$. Left rotors generated by directions orthogonal to the lepton plane are also safe. In particular, the Cabibbo-rung family used in the quark calculation is safe in the local CP chart because its generators lie in $\mathrm{span}\{e_1,e_3\}$.

\subsection{Reality Theorem}\label{subsec:reality-theorem}
The CP Letter gives the full transport-level theorem; for completeness, we record the scalar-product identity needed for the mixing analysis. Consider local lepton representatives of the form
\begin{equation}\label{eq:local-lepton-reps}
    |\ell_a^\pm\rangle=\frac{1}{\sqrt 2}(1\pm i e_a),\qquad e_a\in\Pi_\ell^{\rm loc}.
\end{equation}
Changing the sign in \(|\ell_a^\pm\rangle\) only changes removable diagonal phases and harmless overall signs in the formula below; it does not affect the Jarlskog conclusion. For definiteness, take the plus sign. Expand
\begin{equation}\label{eq:lepton-map-expansion}
    \operatorname{C}(1)=\alpha_0 1+\sum_b\alpha_b e_b,
    \qquad
    \operatorname{C}(e_k)=\beta_0 1+\sum_b\beta_b e_b,
\end{equation}
with all $\alpha_\mu,\beta_\mu\in\mathbb{R}$. A direct scalar-product calculation gives
\begin{align}
    \langle \ell_a^+|\operatorname{C}|\ell_k^+\rangle
    &=\frac{1}{2}\left([\operatorname{C}(1)]_0+[\operatorname{C}(e_k)]_{e_a}\right)\nonumber\\
    &\quad+\frac{i}{2}\left([\operatorname{C}(e_k)]_0-[\operatorname{C}(1)]_{e_a}\right). \label{eq:masterformula}
\end{align}
The imaginary part is precisely the scalar--flavor mixing excluded by the safe condition. A direct derivation of this identity is given in Appendix~\ref{app:leptonformula}. Therefore, if the final effective lepton bridge, including the neutral-sector part, is represented by a safe real-linear map, all such lepton amplitudes are real up to removable phases. In that case,
\begin{equation}\label{eq:leptonCP}
    J_\ell=0,\qquad \delta^\ell_{\rm CP}\in\{0,\pi\}.
\end{equation}
This is a conditional theorem about the final effective operator entering the PMNS amplitudes. It is not a statement about every intermediate factor in a possible bridge decomposition. An intrinsically complex bridge, a real-linear bridge with matrix elements between \(\mathbb C\cdot 1\) and \(\Pi^{\rm loc}_\ell\), the corresponding transported lepton plane in another chart, or neutral-sector bridge structure outside the safe class is the loophole. Such a bridge would generate a genuine leptonic CP phase.

\section{Phenomenological Assessment}\label{sec:phenomenology}

\begin{table*}[htbp]
\centering

\setlength{\tabcolsep}{4pt}
\renewcommand{\arraystretch}{1.9}

\setlength{\colA}{0.20\textwidth}
\setlength{\colB}{0.12\textwidth}
\setlength{\colC}{0.58\textwidth}

\begin{tabular}{lll}
\hline\hline

\begin{minipage}[t]{\colA}
\textbf{Ingredient}
\end{minipage}
&
\begin{minipage}[t]{\colB}
\textbf{Class}
\end{minipage}
&
\begin{minipage}[t]{\colC}
\textbf{Comment}
\end{minipage}
\\[0.4em]
\hline

\begin{minipage}[t]{\colA}
Adjacent-Edge Lift\\
$\tan \theta_{ij}^{(f)}=\sqrt{m_i^{(f)}/m_j^{(f)}}$
\end{minipage}
&
\begin{minipage}[t]{\colB}
Ansatz
\end{minipage}
&
\begin{minipage}[t]{\colC}
\justifying\noindent Exact for the standard Fritzsch two-state texture~\cite{Fritzsch:1977za} in Eq.~\eqref{eq:fritzsch2}; phenomenological when applied as an independent edge rule in the full three-generation problem.
\end{minipage}
\\[0.8em]

\begin{minipage}[t]{\colA}
Cabibbo Rung Phase\\
$\phi_{12}=-2\chi$
\end{minipage}
&
\begin{minipage}[t]{\colB}
Exact
\end{minipage}
&
\begin{minipage}[t]{\colC}
\justifying\noindent
Exact local transport theorem proved in the companion CP Letter~\cite{GuptaTeli:2026aqf}. It supplies the phase available to the Cabibbo block; it is not, by itself, a prediction of the standard CKM Dirac phase.
\end{minipage}
\\[0.8em]

\begin{minipage}[t]{\colA}
Cabibbo Magnitude\\
$|V_{us}|$
\end{minipage}
&
\begin{minipage}[t]{\colB}
Bridge diagnostic
\end{minipage}
&
\begin{minipage}[t]{\colC}
\justifying\noindent
With the fitted companion ratios, the measured \(|V_{us}|\) fixes an effective Cabibbo-block phase \(\phi_{12}\simeq105.7^\circ\) inside the two-angle bridge ansatz. This is not a CKM CP-phase prediction; quadrature is the balanced-rung reference value.
\end{minipage}
\\[0.8em]

\begin{minipage}[t]{\colA}
$(2,3)$ Normalization\\
$\kappa_{23}$
\end{minipage}
&
\begin{minipage}[t]{\colB}
Fitted
\end{minipage}
&
\begin{minipage}[t]{\colC}
\justifying\noindent Phenomenological input fitted to $|V_{cb}|$; with the fitted companion ratios, \(\kappa_{23}\simeq 0.56\). Not yet derived.
\end{minipage}
\\[0.8em]

\begin{minipage}[t]{\colA}
Direct $(1,3)$ Mixing
\end{minipage}
&
\begin{minipage}[t]{\colB}
Open
\end{minipage}
&
\begin{minipage}[t]{\colC}
\justifying\noindent Open long-edge bridge problem; not a parameter-free prediction of the adjacent-edge lift.
\end{minipage}
\\[0.8em]

\begin{minipage}[t]{\colA}
PMNS Angles
\end{minipage}
&
\begin{minipage}[t]{\colB}
Open
\end{minipage}
&
\begin{minipage}[t]{\colC}
\justifying\noindent Not fixed by the neutrino spectrum alone; requires the neutral-sector basis choice or the Higgs bridge.
\end{minipage}
\\[0.8em]

\begin{minipage}[t]{\colA}
Lepton CP Conservation
\end{minipage}
&
\begin{minipage}[t]{\colB}
Conditional
\end{minipage}
&
\begin{minipage}[t]{\colC}
\justifying\noindent Conditional theorem proved in the companion CP Letter~\cite{GuptaTeli:2026aqf} for the final safe real-linear transport; in a transported chart, ``safe'' means no mixing between $\mathbb C\cdot1$ and the corresponding lepton flavor plane.
\end{minipage}
\\[0.8em]

\begin{minipage}[t]{\colA}
Minimal Neutrino Branch
\end{minipage}
&
\begin{minipage}[t]{\colB}
Optional specialization
\end{minipage}
&
\begin{minipage}[t]{\colC}
\justifying\noindent
If the additional perturbative Majorana branch of the companion CP Letter~\cite{GuptaTeli:2026aqf} is imposed, the symmetric spectrum gives inverted ordering and \(m_{\beta\beta}\simeq19\,{\rm meV}\). This branch is quoted here but is not assumed in the general adjacent-edge mixing analysis.
\end{minipage}
\\[0.8em]

\begin{minipage}[t]{\colA}
Higgs Bridge
\end{minipage}
&
\begin{minipage}[t]{\colB}
Open
\end{minipage}
&
\begin{minipage}[t]{\colC}
\justifying\noindent Open dynamical object; its identity--flavor or intrinsically complex components determine whether leptonic CP violation is generated and how the neutral-sector diagonalizer is fixed.
\end{minipage}
\\[0.4em]

\hline\hline
\end{tabular}

\caption{Status of the principal ingredients of the mixing framework.}
\label{tab:status}
\end{table*}

\subsection{Quark Sector}
The quark-sector construction has three layers. First, the mass paper supplies the adjacent square-root ratios. Second, the adjacent-edge lift maps those ratios into local two-generation angles. Third, the octonionic Cabibbo rung supplies the exact relative phase law $\phi_{12}=-2\chi$.

The Cabibbo block is the strongest part of the implementation at the algebraic level: the octonionic rung gives the exact local phase law \(\phi_{12}=-2\chi\). Numerically, using the fitted companion ratios and the minimal two-angle extraction from \(|V_{us}|\) gives the bridge-diagnostic value \(\phi_{12}\simeq 105.7^\circ\). This is an \(\mathcal{O}(1)\) Cabibbo-block phase inside the ansatz, with the balanced rotor at \(\phi_{12}=\pi/2\) serving as the natural geometric reference point rather than an exact fitted value. It should not be read as a prediction of the convention-independent CKM Dirac phase. The \((2,3)\) block is less complete: the direct adjacent-edge estimate gives \(|V_{cb}|_0\simeq0.0751\), so the phenomenological factor \(\kappa_{23}\simeq0.56\) has been introduced. The \((1,3)\) block is not a parameter-free prediction of the adjacent-edge lift; it probes the long-edge bridge.

Thus, the present CKM construction is a controlled phenomenological extension of the mass hierarchy framework, not yet a closed parameter-free derivation of all CKM data. Its derived content is the conjugation relation of the local quark amplitudes and the exact rung phase law. Its fitted or open content is the full three-generation bridge that sets \(\kappa_{23}\), the long-edge suppression, and the map, if any, between the local rung phase and the standard CKM phase.

\subsection{Lepton Sector and Current Data}
The lepton-sector theorem becomes sharply testable only after the safe-bridge assumption is embedded into a concrete lepton bridge. At the general level of the present mixing paper, the charged-lepton edge angles are too small to explain the observed PMNS matrix, and the symmetric neutrino spectrum leaves a basis freedom in the neutral sector. Therefore, the PMNS angles, the neutrino ordering, and the Dirac/Majorana interpretation are not fixed by the adjacent-edge lift alone.

The robust CP statement is conditional. If the final effective lepton bridge is safe, real-linear, then the PMNS amplitudes are real up to removable diagonal phases and
\[
J_\ell=0,\qquad \delta^\ell_{\rm CP}\in\{0,\pi\}.
\]
A confirmed nonzero leptonic Dirac phase would not refute the local transport theorem. It would instead identify the missing bridge structure: scalar--flavor mixing in the real bridge, an intrinsically complex bridge component, or neutral-sector dynamics not representable within the safe real-linear class. Conversely, convergence of the measured phase toward \(0\) or \(\pi\) would keep the safe-bridge explanation viable.

If one additionally imposes the minimal Majorana branch analyzed in the companion CP Letter~\cite{GuptaTeli:2026aqf}, the lepton sector becomes more sharply falsifiable: inverted ordering and \(m_{\beta\beta}\simeq19\,{\rm meV}\) are then part of the same minimal package. At the same time, leptonic CP conservation remains conditional on the absence of the identity--flavor bridge loophole. This experimental package is not rederived in the present paper; it is cited as the CP Letter's minimal specialization. The role of the Mixing paper is to determine where that package fits within the broader CKM/PMNS bridge framework.

Present oscillation data constrain the PMNS angles, the Dirac phase, and the mass ordering~\cite{Esteban:2024fit, NuFIT6.1, T2K:2025wet}. Future long-baseline programs such as DUNE and Hyper-Kamiokande will test whether the safe-bridge CP-conserving limit is viable~\cite{DUNEPhysicsTDR2020, HyperK:2025sensitivity}. In the language of the present framework, a nonzero measured phase would be constructive: it would tell us which bridge component must be added.

\paragraph*{Status of Assumptions.} Table~\ref{tab:status} summarizes the status of the main ingredients. This separation is important because the framework contains exact algebraic statements, fitted spectral moduli, phenomenological bridge parameters, optional minimal specializations, and open dynamical tasks.

\section{Conclusion}\label{sec:conclusion}

\subsection{Summary of Present Results}
We have formulated a companion mixing framework for the exceptional-Jordan spectral approach to fermion masses. The mass paper supplies the ordered Jordan spectral scales and the adjacent square-root mass ratios; the present work asks how far those data can be lifted to CKM and PMNS mixing matrices.

The adjacent-edge lift maps a two-generation square-root mass ratio to a mixing angle. It is exact for a simple Fritzsch-type two-state texture, but it is an ansatz when extended to the full three-generation problem. In the quark sector, the strongest result is the Cabibbo-rung phase law. With the correct octonionic adjoint, the generator \(e^{i\chi}\alpha^\dagger_2-e^{-i\chi}\alpha_2\) is the real octonionic element \(\cos\chi\,e_3-\sin\chi\,e_1\). The corresponding up- and down-sector amplitudes are complex conjugates, leading to the exact local law \(\phi_{12}=-2\chi\). With the fitted companion ratios, the minimal two-angle extraction from the measured Cabibbo magnitude gives the bridge-diagnostic value \(\phi_{12}\simeq 105.7^\circ\), while the balanced rotor gives the distinguished quadrature reference value. This is not a prediction of the standard CKM Dirac phase. The remaining CKM sectors are not yet fully derived: \(|V_{cb}|\) requires the effective normalization \(\kappa_{23}\simeq0.56\), \(|V_{ub}|\) probes a suppressed long-edge bridge, and the map from the local rung phase to the Jarlskog phase requires the full weak-to-mass bridge.

In the lepton sector, the charged-lepton edge rotations are small compared with the observed PMNS angles. The large PMNS mixing must therefore come from the neutral-sector diagonalizer, the Higgs bridge, or both. Because the symmetric neutrino Jordan spectrum contains a degenerate pair in magnitude, the mass eigenvalues alone do not determine \(U_{\nu, L}\). At the level of the general mixing framework, this leaves the ordering and the Dirac/Majorana interpretation to the neutral-sector bridge. The companion CP Letter's additional minimal Majorana branch gives a sharper specialization, with inverted ordering and \(m_{\beta\beta}\simeq19\,{\rm meV}\), but that branch is not an input to the adjacent-edge lift.

The CP statement used here is correspondingly conditional: if the final effective lepton bridge, including the neutral-sector part, is safe and real-linear, all lepton amplitudes are real up to removable phases and \(\delta^\ell_{\rm CP}\in\{0,\pi\}\). A confirmed nonzero leptonic CP phase would therefore identify the required loophole: scalar--flavor mixing, an intrinsically complex bridge, or neutral-sector bridge dynamics outside the safe real-linear class.

The framework, therefore, separates four physical categories. First, the pure transport-level consequences of the octonionic Clifford structure yield rigid local results, most notably the exact Cabibbo-rung phase law and the conditional lepton reality theorem, both proved in the companion CP Letter~\cite{GuptaTeli:2026aqf}. Second, the fitted spectral moduli of the mass paper supply the adjacent hierarchy ratios used as inputs to the mixing ansatz. Third, the CP Letter's minimal Majorana branch is an optional specialization of the neutral sector, not part of the general adjacent-edge lift. Fourth, the open dynamical task is the weak-to-mass bridge: it must explain \(\kappa_{23}\), the long-edge suppression in \(V_{ub}\), the neutral-sector basis choice, the relation between the local rung phase and the standard CKM phase, and any scalar--flavor component responsible for leptonic CP violation.

\subsection{Dynamical Outlook: Trace Dynamics and Exceptional Geometry}
While the algebraic constraints of the exceptional-Jordan spectral framework provide a tight kinematic starting point, a complete description requires moving beyond the purely spectral setup. The open dynamical tasks isolated in this work, calculating the suppressed long-edge \((1,3)\) quark mixing, deriving the \((2,3)\) normalization, fixing the neutral-sector basis, determining the relation between the local Cabibbo-rung phase and the standard CKM phase, and determining the identity--flavor content of the Higgs bridge point toward a deeper pre-quantum matrix layer.

A natural candidate for this underlying framework is Adlerian trace dynamics~\cite{Adler:2023hdu}, together with its noncommutative and octonionic extensions~\cite{Lochan:2011nr, Roy:2020rlw, Singh:2024ven}. In such a framework, the adjacent-edge lift ansatz and the bridge matrix elements would not be inserted as empirical inputs but would instead arise as effective statistical or dynamical data from an underlying trace action. Recent developments also suggest structural links between generalized trace dynamics and generalized noncommutative geometries~\cite{Farnsworth:2026aae}. The immediate dynamical target is therefore concrete: compute the weak-to-mass bridge in the same algebraic setting, and determine whether the open quantities isolated in Table~\ref{tab:status} become predictions or explicit failure modes.

\begin{acknowledgments}
B.\,G.\,T. would like to thank Dawood Kothawala, Department of Physics, IIT Madras, for his guidance and support as co-supervisor, and for enabling this work to be partly carried out as part of the M.Sc. project. He also thanks the Inter-University Center for Astronomy and Astrophysics (IUCAA), Pune, for hospitality and accommodation during a research visit.
\end{acknowledgments}

\appendix

\section{Proof of the Lepton Reality Theorem}\label{app:leptonformula}
For completeness, we derive the scalar-product identity used in Sec.~\ref{subsec:reality-theorem}. The full transport-level formulation is given in the companion CP Letter~\cite{GuptaTeli:2026aqf}. Let $\operatorname{C}:\mathbb{O}_\mathbb{C} \to \mathbb{O}_\mathbb{C}$ be real-linear in the sense of Sec.~\ref{subsec:lepton-safe-transports} and write
\begin{equation}\label{eq:appendix-lepton-map-expansion}
    \operatorname{C}(1)=\alpha_0 1+\sum_b\alpha_b e_b,
    \qquad
    \operatorname{C}(e_k)=\beta_0 1+\sum_b\beta_b e_b,
\end{equation}
with real coefficients. For
\[
|\ell^+_a\rangle=\frac{1+i e_a}{\sqrt2},
\]
the combined complex-plus-octonionic adjoint gives
\((1+i e_a)^\dagger=(1+i e_a)\). With the inner-product convention
used here,
\begin{align}
\langle \ell^+_a|\operatorname{C}|\ell^+_k\rangle
&=
\frac12\,\operatorname{Sc}\!\left[(1+i e_a)\operatorname{C}(1+i e_k)\right]
\nonumber\\
&=
\frac12\,\operatorname{Sc}\!\left[(1+i e_a)\big(\operatorname{C}(1)+i\operatorname{C}(e_k)\big)\right].
\label{eq:lepton-master-intermediate}
\end{align}
Using
\[
\operatorname{Sc}(1)=1,
\qquad
\operatorname{Sc}(e_b)=0,
\qquad
\operatorname{Sc}(e_a e_b)=-\delta_{ab},
\]
all terms containing distinct octonionic imaginary units drop out under scalar projection, giving
\begin{equation}\label{eq:A3}
    \langle \ell_a^+|\operatorname{C}|\ell_k^+\rangle
    =\frac{1}{2}\left(\alpha_0+\beta_a\right)
    +\frac{i}{2}\left(\beta_0-\alpha_a\right),
\end{equation}
which is Eq.~\eqref{eq:masterformula}. Hence
\[
\Im\langle \ell_a^+|\operatorname{C}|\ell_k^+\rangle=\frac12(\beta_0-\alpha_a)
\]
is precisely the scalar--flavor mixing term. Hence Eq.~\eqref{eq:A3} shows that the safe condition is sufficient for the reality of the amplitudes. Within the rotor/channel transport class considered in the text, it is also the relevant condition for excluding physical lepton phases. For a completely general real-linear map, accidental cancellations with $\beta_0=\alpha_a$ are possible. This establishes the conditional reality theorem stated in Sec.~\ref{subsec:reality-theorem}.

\bibliography{references}

\end{document}